\documentclass[doublecol]{epl2}
\usepackage{pdflscape}
\usepackage{amsmath}
\usepackage{amssymb}

\newcommand{\rL}{\rho_\Lambda}

\newcommand{\CC}{\Lambda}

\newcommand{\rmr}{\rho_m}

\newcommand{\be}{\begin{equation}}
\newcommand{\ee}{\end{equation}}
\def\beq{\begin{equation}}
\def\eeq{\end{equation}}
\def\ber{\begin{eqnarray}}
\def\eer{\end{eqnarray}}

\title{Relaxing the $\sigma_8$-tension through running vacuum in the Universe}

\author{Adri\`a G\'omez-Valent\ 
and  Joan Sol\`a}

\institute{
Departament de F\'isica Qu\`antica i Astrof\'isica, and Institute of Cosmos Sciences, Universitat de Barcelona, \\ Av. Diagonal 647, E-08028 Barcelona, Catalonia, Spain}


\abstract{It has recently been shown that the class of running vacuum models (RVMs)
has the capacity to fit the overall cosmological observations better than the concordance $\CC$CDM model, therefore supporting the possibility of dynamical dark energy (DE).
Apart from the cosmic microwave background (CMB) anisotropies, the most crucial datasets  involved  are: i) baryonic acoustic oscillations (BAO), and ii) direct large scale structure (LSS) formation data. Analyses mainly focusing on CMB and with insufficient BAO+LSS input  generally fail to capture the dynamical DE signature, whereas the few existing studies accounting for the wealth of known CMB+BAO+LSS data (see in particular  Sol\`a, G\'omez-Valent \& de Cruz P\'erez 2015, 2017; and  Zhao et al. 2017) do converge to the remarkable conclusion that dynamical DE might well be encoded in the current cosmological observations at  $3-4\sigma$ c.l. A decisive factor  is  the persistent $\sigma_8$-tension between the $\CC$CDM and the data.
Because the issue is obviously pressing, we devote this work to explain  how and why running vacuum in the expanding universe successfully  relaxes the existing $\sigma_8$-tension and describes the LSS  formation data significantly better than the  $\CC$CDM.}

\begin{document}

\maketitle

\section{Introduction}
\label{intro}

The concordance cosmological model, or standard $\CC$CDM model of cosmology \cite{Peebles1984,Peebles1993}, is generally considered the most successful description of the overall cosmological data known to date. A bounty of modern observations of different kinds speaks up in favor of this fact (cf. \cite{Planck2016,DES2017}). A principal building block of the $\CC$CDM   is the cosmological term  $\CC$, of which we celebrate this year a century of its existence in Einstein's equations \cite{Einstein1917}. It has traditionally been associated to the concept of vacuum energy density, $\rL = \CC/(8\pi G)$  ($G$ being Newton's constant), and is thought to be the cause for the  accelerated expansion of our universe. Such speeding up of our cosmos was discovered almost twenty years ago from the measurement of the apparent magnitude versus redshift relation of distant Supernovae of type Ia (SNIa) \cite{Riess1998,Perlmutter1999}. Despite the initial phenomenological success, at a more fundamental ground quantum field theory (QFT) establishes a connection between $\rL$ and the quantum vacuum which is at the root of the famous (so far unsolved) cosmological constant (CC) problem \cite{Weinberg1989,Padmanabhan2003,PeeblesRatra2003,SolaReview2013}. The various faces of it (including the so-called cosmic coincidence problem)  indicate  that the idea of a strictly constant $\CC$ could be an oversimplification, even at the pure phenomenological level. That this may well be the case is suggested by the fact that other, more recent, observational pitfalls have been pestering the  straightforward viability of the $\CC$CDM.  Among them the so-called $H_0$ and $\sigma_8$-tensions, showing significant and persistent discrepancies of the standard model prediction with the cosmological observations.

\begin{table*}
\begin{center}
\resizebox{1\textwidth}{!}{
\begin{tabular}{ |c|c|c|c|c|c|c|c|c|c|}
\multicolumn{1}{c}{Model} &  \multicolumn{1}{c}{$H_0$(km/s/Mpc)} &  \multicolumn{1}{c}{$\omega_b$} & \multicolumn{1}{c}{{\small$n_s$}}  &  \multicolumn{1}{c}{$\Omega_m^0$} &\multicolumn{1}{c}{$\nu_i$} &\multicolumn{1}{c}{$w$} &\multicolumn{1}{c}{$\chi^2_{\rm min}/dof$} & \multicolumn{1}{c}{$\Delta{\rm AIC}$} & \multicolumn{1}{c}{$\Delta{\rm BIC}$}\vspace{0.5mm}
\\\hline
$\Lambda$CDM  & $68.83\pm 0.34$ & $0.02243\pm 0.00013$ &$0.973\pm 0.004$& $0.298\pm 0.004$ & - & -1  & 84.40/85 & - & - \\
\hline
XCDM  & $67.16\pm 0.67$& $0.02251\pm0.00013 $&$0.975\pm0.004$& $0.311\pm0.006$ & - &$-0.936\pm{0.023}$  & 76.80/84 & 5.35 & 3.11 \\
\hline
RVM  & $67.45\pm 0.48$& $0.02224\pm0.00014 $&$0.964\pm0.004$& $0.304\pm0.005$ &$0.00158\pm 0.00041 $ & -1  & 68.67/84 & 13.48 & 11.24 \\
\hline
\end{tabular}}
\caption{{ Best-fit values obtained from the SNIa+BAO+$H(z)$+LSS+CMB fitting analysis of \cite{SoGoPePLB2017} for the $\CC$CDM, XCDM, and the RVM. See the aforementioned paper for additional information about the statistical significance of the results. Also for the complete list of data used in the analysis and the corresponding references. Both, the XCDM and the RVM are clearly preferred over the $\Lambda$CDM. The positive signal in favor of vacuum dynamics reaches  $\sim 3.8\sigma$ c.l. in the RVM. In the XCDM parametrization the dynamical DE signal is lower, but still quite high. It reaches  $\sim 2.8\sigma$ c.l. The Akaike and Bayesian  criteria in the last two columns (see text) do reconfirm these important signs of dynamical DE.}}
\end{center}
\label{tableFit1}
\end{table*}

%

 There is indeed currently a notable tension between CMB measurements of $H_0$ \cite{Planck2016} and local (distant ladder) determinations \cite{Riess2016}.  At the same time there is an ongoing tension in the large scale structure (LSS) formation data, which is described in terms of the combined observable $f(z)\sigma_8(z)$, where $f(z)$ is the linear growth rate and $\sigma_8$  the RMS matter fluctuation on scales of $R_8 = 8{h^{-1}}$ Mpc\,\cite{Planck2016}. The corresponding prediction of the $\CC$CDM is known to be too large and the overall description of the LSS data points is not well accounted for by the concordance model\,\cite{Macaulay2013}. Such preference for  lower  $\sigma_8$ values appears to agree with recent measurements, see e.g.\,\cite{Joudaki2017,Henning2017,Hildebrandt2017}.  In \cite{SoGoPeIJMP2017,SoGoPePLB2017} a detailed study was made on the $H_0$-tension in connection to vacuum dynamics.  In the present Letter we will  instead focus on the $\sigma_8$-tension, and shall show that it can be fully relaxed in the context of the class of running vacuum models (RVMs), in which $\rL$ is no longer a rigid quantity but a slow dynamical variable evolving with the Hubble rate, $H$, i.e.  $\rL(H)$.

Our approach at solving  the $\sigma_8$-tension, however, is not just to cure an isolated conflict in the LSS or  CMB+LSS sectors, but to derive it  harmonically from an overall fit to the cosmological data SNIa+BAO+$H(z)$+LSS+CMB. Recently, it has been demonstrated that the RVMs yield a global fit to cosmological  data which is better than that of the  $\CC$CDM at $3-4\sigma$ c.l. The first significant signs were reported in  \cite{SoGoPeApJ2017,SoGoPeApJL2015}, with subsequent analyses in\,\cite{SoGoPeMPLA2017,SoGoPeIJMP2017,SoGoPePLB2017}. Another recent and fully independent study of dynamical DE, based on similar datasets and nonparametric methods,\,\cite{Wang2015},  reaches a similar conclusion\, \cite{GongBoZhao2017}.

We devote this Letter to show and explain why the running vacuum models do completely relax the $\sigma_8$-tension and  provide an improved  description of the LSS  formation data as compared to the rigid vacuum option offered by the traditional $\CC$CDM model. Such explanation might well be the touchstone for definitely  pinpointing significant signs of dynamical DE in the expanding universe.

\section{Running vacuum in interaction with matter}

The RVM is a dynamical vacuum model, meaning that its  equation of state (EoS) parameter is still $w=-1$ but the  vacuum energy density is a ``running'' one, i.e. it evolves with expansion, but departs  (mildly) from the rigid assumption $\rL=$const. of the $\CC$CDM. Specifically, the form of $\rL$ in the RVM case reads  (for reviews, see\,\cite{SolaReview2013,Sola2011,SoGo2015,SolaReview2016,GomezValentPhD} and references therein):
\begin{equation}\label{eq:RVMvacuumdadensity}
\rho_\CC(H) = \frac{3}{8\pi{G}}\left(c_{0} + \nu{H^2}\right)\,.
\end{equation}
Here $c_0=H_0^2\left(1-\Omega_m-\Omega_r-\nu\right)$  is determined  by the boundary condition $\rL(H_0)=\rho_{\Lambda 0}=\rho_{c0}\,(1-\Omega_m-\Omega_r)$, with $\rho_{c0}=3H_0^2/(8\pi G)$ the critical density, and $\Omega_m=\Omega_b+\Omega_{dm}$ the matter density parameter at present, being the sum of the baryon part and the dark matter (DM) contribution. The dimensionless coefficient $\nu$ is the pivotal vacuum parameter under study, whose nonvanishing value is responsible for the cosmic evolution of the vacuum energy density in the RVM.  It is expected to be very small, $|\nu|\ll1$, since the model must remain sufficiently close to the $\CC$CDM. The moderate dynamical evolution of $\rL(H)$ is possible thanks to the mild vacuum-matter interaction (see below). Formally $\nu$ can be given a QFT meaning by linking it to the $\beta$-function of the running $\rL$ (see \cite{SolaReview2013,Sola2015}). Theoretical estimates place its value in the ballpark of $\nu\sim 10^{-3}$ at most in the context of a typical Grand Unified Theory (GUT) \cite{Sola2008}. Remarkably, this is precisely the order of magnitude derived from extensive analyses of the cosmological data \cite{GoSoBa2015,SoGoPeApJL2015,SoGoPeIJMP2017,SoGoPePLB2017,SoGoPeApJ2017,SoGoPeMPLA2017,SoPeGoPRL2016,SoPeGoPRD2017}.


\begin{figure*}
\begin{center}
\label{FigLSS1}
\includegraphics[width=5.0in]{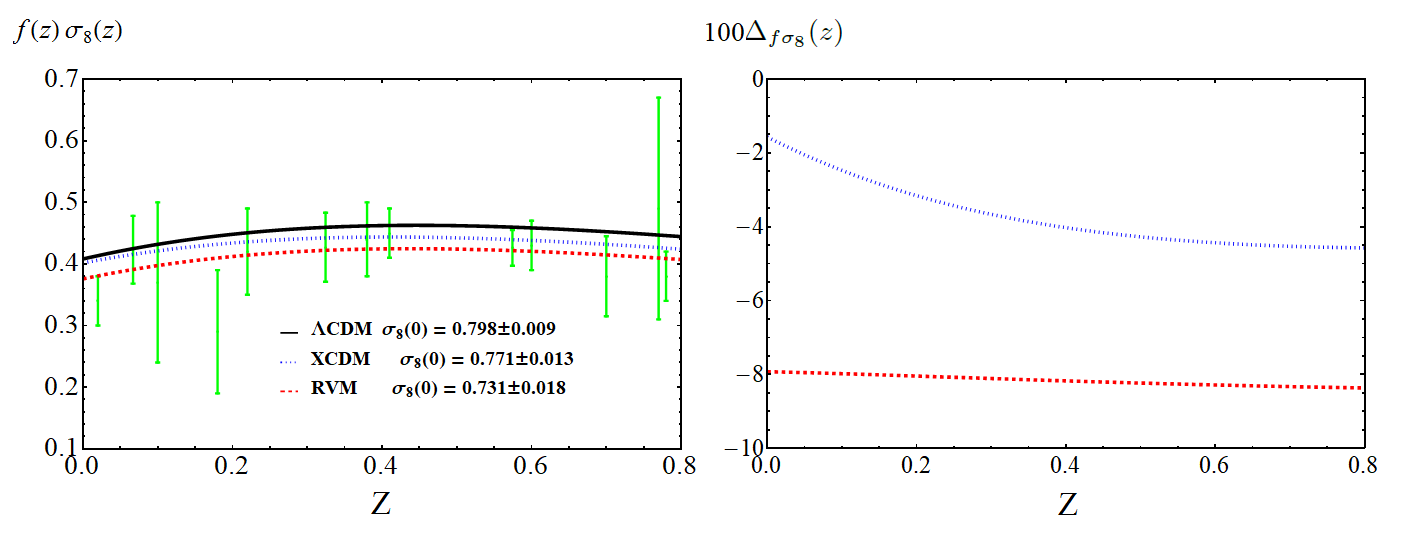}\ \ 
\caption{{\it Left:} The $f(z)\sigma_8(z)$ data points and the theoretical curves for  $\Lambda$CDM,  XCDM and  RVM. The values of $\sigma_8$ that we obtain for these models are quoted in the figure; {\it Right:} The relative difference in the theoretical curves  for the RVM and XCDM with respect to $\Lambda$CDM,  $\Delta_{f\sigma_8}(z)$ (in $\%$). In the RVM  case it is computed from Eq.\,\eqref{eq:Dfs8}, and similarly for the XCDM.}
\end{center}
\end{figure*}


Different realizations of the RVM are possible, but here we limit ourselves to the canonical option in which  only the DM  is exchanging energy with the vacuum and hence  under covariant conservation of radiation and baryons, i.e. their energy densities evolve in the standard way: $\rho_r(a)=\rho_{r0}\,a^{-4}$ and $\rho_b(a) = \rho_{b0}\,a^{-3}$. The dynamical evolution of  $\rL$ can therefore exclusively be associated to the exchange of energy with the DM. The coupled system of conservation equations is
\begin{eqnarray}
&&\dot{\rho}_{dm}+3H\rho_{dm}=Q\label{eq:Qequations1}\\
&&\dot{\rho}_\CC =-Q\,.\label{eq:Qequations2}
\end{eqnarray}
 With $\rL(H)$ given by (\ref{eq:RVMvacuumdadensity}),  the Friedmann and acceleration equations take on the same form as in the standard case:
\begin{eqnarray}
&&3H^2=8\pi\,G\,(\rho_m+\rho_r+\rho_\Lambda(H))\label{eq:FriedmannEq}\\
&&3H^2+2\dot{H}=-8\pi\,G\,(p_r + p_\Lambda(H))\,.\label{eq:PressureEq}
\end{eqnarray}
Here $H=\dot{a}/a$ is the Hubble function, $a(t)$ the scale factor as a function of the cosmic time and $p_r=\rho_r/3$ is the radiation pressure.  Combining the above two equations, we get  $\dot{H}=-(4\pi G/3)\,\left(3\rho_m+4\rho_r\right)$,
and upon differentiating (\ref{eq:RVMvacuumdadensity}) with respect to the cosmic time we are led to
$\dot{\rho}_\CC=-\nu\,H\left(3\rho_m+4\rho_r\right)$.
Thus, we find
\begin{equation}\label{eq:QforModelRVM}
 Q=\nu\,H(3\rho_{m}+4\rho_r)=\nu\,H(3\rho_{b}+3\rho_{dm}+4\rho_r)
\end{equation}
for the source function $Q$ in equations (\ref{eq:Qequations1}) and (\ref{eq:Qequations2}). We see that, contrary to other phenomenological models in the literature (discussed e.g. in \cite{SoPeGoPRD2017}), the form of $Q$ is precisely determined by the form of the vacuum energy density (\ref{eq:RVMvacuumdadensity}), which on its own is motivated in the QFT context as indicated above.

Inserting Eq.\,(\ref{eq:QforModelRVM}) into equations (\ref{eq:Qequations1}) and (\ref{eq:Qequations2}) and integrating the resulting differential equations in terms of the scale factor, we find for the total matter density $\rho_m= \rho_{dm}+ \rho_{b}$  ({see \cite{SoPeGoPRD2017} for details}):
\begin{eqnarray}
\rho_{m}(a) = \rho_{m0}\,a^{-3(1-\nu)}
+\frac{4\nu}{1 + 3\nu}\, \rho_{r0}\left(a^{-3(1-\nu)} - a^{-4}\right)\,. \label{eq:rhoRVM}
\end{eqnarray}
 The Hubble function is found to be
\begin{eqnarray}\label{eq:E2RVM}
E^2(a) &=& 1 + \frac{\Omega_m}{1-\nu}\left(a^{-3(1-\nu)}-1\right) \label{HRVM}\\
&& + \frac{\Omega_r}{1-\nu}\left(\frac{1-\nu}{1+3\nu}a^{-4} + \frac{4\nu}{1+3\nu}a^{-3(1-\nu)} -1\right)\,.\nonumber
\end{eqnarray}
Here  $E\equiv H/H_0$ is the normalized Hubble rate with respect to the current value. One can easily check that for $\nu=0$ we recover the $\CC$CDM expressions, as expected. We note that $E(1)=1$, as also expected.
%
%
%
As a baseline reference model, it is also convenient to fit  the same data through the simple XCDM parametrization of the dynamical DE \cite{TurnerWhite1997}. Since both matter and DE are self-conserved in the XCDM (i.e., they are not interacting), the DE density as a function of the scale factor is simply given by $\rho_X(a)=\rho_{X0}\,a^{-3(1+w)}$, with $\rho_{X0}=\rho_{\CC 0}$, where $w$ is the (constant) EoS parameter of  the generic DE entity X.
For $w=-1$ it boils down, of course, to that of the $\CC$CDM with rigid $\CC$-term. The XCDM parametrization is also useful to  mimic a (noninteractive) DE scalar field with constant EoS. For $w\gtrsim-1$ the XCDM mimics quintessence, whereas for $w\lesssim-1$ it mimics phantom DE.

\section{The role of the LSS data}
The large scale structure formation data play a momentous role in the diagnostic of DE. For the $\CC$CDM and XCDM  the matter perturbations are described with the standard equation \cite{Peebles1993}
\begin{equation}\label{diffeqLCDM}
\ddot{\delta}_m+2H\,\dot{\delta}_m-4\pi
G\rmr\,\delta_m=0\,,
\end{equation}
using the corresponding  Hubble function for each one of these models.
In the RVM case the perturbations equation itself becomes modified. It has been studied in detail e.g. in \cite{GoSoBa2015} and \cite{SoPeGoPRD2017}. It reads:
\begin{equation}\label{diffeqD}
\ddot{\delta}_m+\left(2H+\Psi\right)\,\dot{\delta}_m-\left(4\pi
G\rmr-2H\Psi-\dot{\Psi}\right)\,\delta_m=0\,,
\end{equation}
where $\Psi\equiv -\dot{\rho}_{\Lambda}/{\rmr}= Q/{\rmr}$, and the interaction source $Q$ is given by (\ref{eq:QforModelRVM}). For $\nu=0$ (hence  $Q=0$ ) equation \eqref{diffeqD} reduces to \eqref{diffeqLCDM}.
We note that at the scales under consideration (subhorizon) we are neglecting the perturbations of the vacuum energy density in front of the perturbations of the matter field. This can be substantiated in different ways, see e.g. \cite{GranPelSol2009} and the above mentioned references from which (\ref{diffeqD}) is derived. For a discussion in two different gauges,  conformal Newtonian and synchronous \cite{MaBert1995}, see  the detailed study \cite{GoSoPert}.
 Let us also mention that  one can always choose a frame in which the vacuum is spatially homogeneous using  the comoving synchronous gauge \cite{Wang2013}
To solve the above perturbations equations we have to fix the initial conditions on $\delta_m$ and its derivative for each model at high redshift, namely when non-relativistic matter dominates over radiation and DE. The concrete procedure is explained e.g. in\,\cite{SoPeGoPRD2017}.


\begin{figure*}
\begin{center}
\includegraphics[scale=0.72]{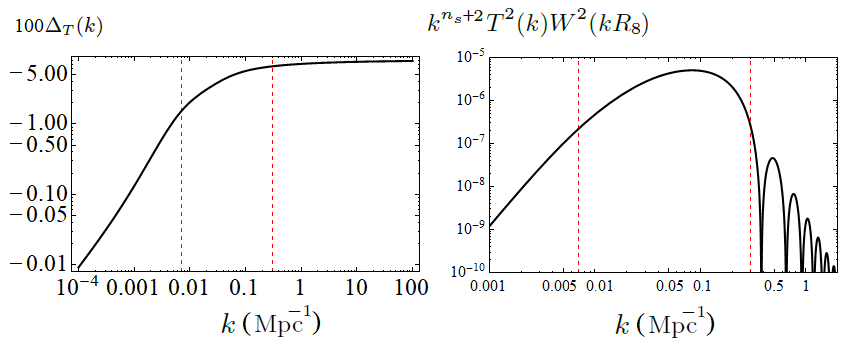}
\caption{{\it Left:} Relative difference  (in $\%$)  between the transfer function \eqref{eq:BBKS} of the RVM (parameters as in Table 1) and of the $\Lambda$CDM (with $\nu=0$ but the other parameters as before) as a function of $k$, i.e. $\Delta_T(k)= (T_{\rm RVM}(k)-T_\Lambda(k))/T_{\Lambda}(k)$; {\it Right:} Behavior of the function in the integrand  of Eq.\,\eqref{eq:s88general}. The range of wave numbers $k$ at which such function is significant for generating the relative differences induced by  $\nu$ on $T(k)$  is marked off with red vertical dashed lines in both plots.
\label{fig:RVMdifOrigin4}}
\end{center}
\end{figure*}


The analysis of the linear LSS data is usually accomplished with the help of the weighted linear growth $f(z)\sigma_8(z)$, where $f(z)=d\ln{\delta_m}/d\ln{a}$ is the growth factor and $\sigma_8(z)$ is the RMS mass fluctuation on $R_8=8\,h^{-1}$ Mpc spheres. The latter  is computed as follows:
\begin{equation}
\begin{small}\sigma_{\rm 8}(z)=\sigma_{8, \CC}
\frac{\delta_m(z)}{\delta^{\CC}_{m}(0)}
\sqrt{\frac{\int_{0}^{\infty} k^{n_s+2} T^{2}(\vec{p},k)
W^2(kR_{8}) dk} {\int_{0}^{\infty} k^{n_{s,\CC}+2} T^{2}(\vec{p}_\Lambda,k) W^2(kR_{8,\Lambda}) dk}}\,,\label{eq:s88general}
\end{small}\end{equation}
where $W$ is a top-hat smoothing function and $T(\vec{p},k)$ the matter transfer function.   We have adopted the usual BBKS form \cite{BBKS1986}:
\begin{equation}\label{eq:BBKS}
\begin{array}{ll}
T(x) = &\frac{\ln (1+0.171 x)}{0.171\,x}\Big[1+0.284 x + (1.18 x)^2+\\
   & + \, (0.399 x)^3+(0.490x)^4\Big]^{-1/4}\,.
\end{array}
\end{equation}
Here  we have defined
$x=k/(k_{eq}\tilde{\Gamma})$, with $k_{eq}=a_{eq}H(a_{eq})$   the value of the comoving wave number at the equality scale $a_{eq}$ between matter and radiation densities, and $\tilde{\Gamma}=
\exp{\{-\Omega_b-\sqrt{2h}\frac{\Omega_b}{\Omega_m}\}}$ is the modified shape parameter \cite{PeacockDodds1994,Sugiyama1995,HuSugiyama1995}.
The fitting parameters for each model are contained in $\vec{p}$ of \eqref{eq:s88general}.
Following \cite{SoGoPeApJ2017,SoPeGoPRD2017,SoGoPePLB2017}, we have taken as fiducial model the $\CC$CDM at fixed parameter values from the Planck 2015 TT,TE,EE+lowP+lensing data \cite{Planck2016}. These fiducial values are collected in $\vec{p}_\CC$. {We have checked that using the Eisenstein \& Hu transfer function\,\cite{EisensteinHu1998} we obtain the same results.}

\section{Extracting the dynamical DE signal}

Recall that we do not wish to discuss the $\sigma_8$-tension as an isolated problem. We consider it in the context of an overall fit to the rich data string SNIa+BAO+$H(z)$+LSS+CMB discussed in detail in \cite{SoGoPeApJ2017,SoPeGoPRD2017,SoGoPePLB2017}. The results are indicated in Table 1. The corresponding  $\chi^2_{\rm min}$ values for the three models under consideration ($\CC$CDM, XCDM and RVM) are quoted in that table.  Clearly the two dynamical DE models perform better than the $\CC$CDM since their $\chi^2_{\rm min}$  values are smaller, and the RVM is the preferred one.  However, the dynamical DE models have one additional parameter as compared to the $\CC$CDM and therefore the comparison is uneven.
For a fairer assessment of the fit quality it proves useful to invoke the time-honored  Akaike and Bayesian information criteria, AIC and BIC, defined as \cite{KassRaftery1995}:
${\rm AIC}=\chi^2_{\rm min}+2nN/(N-n-1)$ and ${\rm BIC}=\chi^2_{\rm min}+n\,\ln N$, where $n$ is the number of fitting parameters and $N$ the number of data points. In this way the competing models with more parameters used to analyze the same data receive a  suitable penalty. The bigger are the differences $\Delta$AIC ($\Delta$BIC) with respect to the model with smaller value of AIC (BIC) -- the XCDM and RVM here -- the higher is the evidence against the model with larger value of  AIC (BIC) -- the $\CC$CDM, in this case. From Table 1 we may compute $\ln A\equiv \Delta{\rm AIC}/2$ and $\ln B\equiv \Delta{\rm BIC}/2$. They represent the Akaike and Bayesian evidences (e.g. $B$ is the Bayes factor, giving the ratio of marginal likelihoods between the two models).
Positive (negative) values of  $\ln A$ and $\ln B$  advocate  in favor (against) dynamical DE. For values above $+5$  one can speak of ``very strong'' evidence against the $\CC$CDM, and hence in support of dynamical DE \cite{KassRaftery1995}.
It follows from Table 1 that the RVM  is greatly preferred over the $\CC$CDM, as $\ln A$ and $\ln B$ are both above $+5$. The preference for the XCDM over the $\CC$CDM is also high, but below that of RVM.

\begin{figure*}
\begin{center}
\label{FigLSS2}
\includegraphics[width=4.5in]{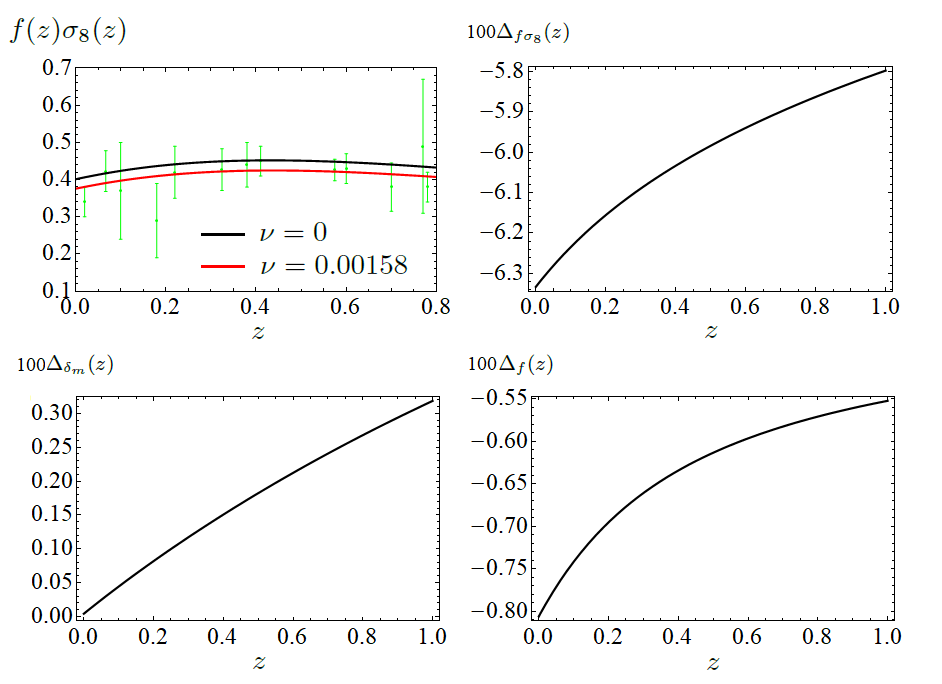}
\caption{{\it Upper-left:} Weighted growth rate obtained by (i) setting all the parameters to the RVM ones (cf. Table 1); and (ii) keeping the same setting,  except $\nu=0$. These correspond to the red and black lines, respectively; {\it Upper-right:} Relative difference $\Delta_{f\sigma_8}(z)$  (in $\%$)  between the curves of the upper-left plot, see Eq.\,(\ref{eq:Dfs8}). The change induced by the nonvanishing vacuum parameter $\nu$ reaches about  $- 6.3\%$ at $z\sim 0$; {\it Lower-left:} Relative difference (in $\%$)  between the density contrasts $\delta_m(z)$  associated to the two scenarios explored in the upper-left plot. The differences in this case are positive and lower than $0.4\%$ for $z<1$; {\it Lower-right:} As before, but for the growth function $f(z)$.  Around the present time, $\Delta_{f}(0)\simeq-0.8\%$.
}
\end{center}
\end{figure*}


A very significant data set with nontrivial  impact on the fit quality of the competing cosmological models is the set of LSS  data points:  $f(z)\sigma_8(z)$.  In Fig. 1 (left) we display  $f(z)\sigma_8(z)$ for the RVM and XCDM models with the fitted values indicated in  Table 1.  We can see that the predicted curve of $f(z)\sigma_8(z)$ for the $\CC$CDM is too high as compared to most of the data, and as a result the global fit quality of the standard model becomes damaged. In stark contrast, both the XCDM and specially the RVM provide an enhanced fit since the corresponding curves are lowered with respect to the $\CC$CDM one. The $\sim 8\%$  negative correction (cf. Fig. 1 right) provided by the RVM is just in the right ballpark to enable theory to predict an excellent agreement with the data.  Not surprisingly the RVM fit quality to the overall data string (SNIa+BAO+$H(z)$+LSS+CMB)  is neatly superior to that of the $\CC$CDM  (cf. Table 1). The XCDM fit, on the other hand, lies in an intermediate position over the  $\CC$CDM, showing that dynamical DE is generically favored by the LSS data.  We may ask ourselves  how it comes about that the RVM is such a particularly gifted option for dynamical DE, or put another way, how is it possible that the small parameter $\nu\sim 10^{-3}$ (cf. Table 1) can eventually trigger a $\sim 8\%$ reduction of the weighted growth rate $f(z)\sigma_8(z)$  thus hitting the bull's-eye of the LSS data much better than the $\CC$CDM? The answer is given in the next section.

\section{A natural solution to the $\sigma_8$-tension}

The mass variance of the smoothed linear density field on scales of  $8 h^{-1}$ Mpc at redshift $z$  is given by \eqref{eq:s88general}.  Its value at $z=0$  is called  $\sigma_8$ and it is a frequent substitute for the parameter $A_s$ normalizing the power spectrum (cf. \cite{Planck2016,DES2017}). It is well-known that the  $\CC$CDM predicts a value of $\sigma_8$ which is too  large (cf. Fig. 1) and hence it leads to an exceeding structure formation power which is unable to explain the LSS observations, represented by the data points on the combined observable  $f(z)\sigma_8(z)$, {see e.g. \cite{BasilakosNesseris2017}  and references therein. For alternative approaches, in which effects of viscosity or of a small amount of spatial curvature are included, see e.g. \cite{Anand2017,Ooba2017ab}}. The reason for fitting the growth rate  $f(z)$ weighted by the value of  $\sigma_8(z)$ at each redshift rather than  $f(z)$  individually  is because the product  $f(z)\sigma_8(z)$ is  independent of the bias factor between the observed galaxy spectrum and the underlying (total) matter power spectrum  \cite{SongPercival2009}. As can be seen from Fig.\,1, the function  $f(z)\sigma_8(z)$ is roughly flat in the redshift range $z\in[0,0.8]$ for the models under study and its overall behavior is essentially determined by its value at $z=0$. Let us focus on  $\sigma_8\equiv\sigma_8(0)$, which plays a particularly relevant role in this story.  From Fig. 1 we can read $\sigma_8 (\CC {\rm CDM})=0.798\pm 0.009$ for the $\CC$CDM model, whereas the RVM prediction is $\sigma_8({\rm RVM})=0.731\pm 0.018$. The central value of $\sigma_8$ for the RVM  is $8.4\%$ smaller than that of the $\CC$CDM. The difference is significant since the two values of $\sigma_8$  differ by $3.3\sigma$ and such discrepancy is the touchstone to explain the improvement of the RVM fit with respect to the $\CC$CDM one. Another factor  explaining such improvement is the superior description of BAO data by the  RVM \cite{SoPeGoPRD2017}, but here we shall concentrate on the LSS data, as such data are responsible for the acute $\sigma_8$-tension observed between the $\CC$CDM prediction and the observations.

 In the following we will show analytically and numerically how the vacuum coefficient $\nu$ of the RVM is able to fully relax such tension comfortably, without transferring it to any other parameter of the model, and hence provides  the  necessary reduction in the value of $\sigma_8$   -- and consequently of   $f(z)\sigma_8(z)$, see Fig. 1 -- with respect to the $\CC$CDM. We want to show that the main effect comes from the correction introduced by $\nu\neq0$ on the transfer function  (\ref{eq:BBKS}).
 Let us first determine the ($\nu$-dependent) equality scale  between matter and radiation densities in the RVM: $\rho_r(a_{eq})=\rho_m(a_{eq})$, in which $\rho_m(a)$ is the function (\ref{eq:rhoRVM}) and $\rho_r(a)$ is the standard one for conserved radiation.  We find
 \begin{equation}\label{eq:aeqRVM}
 a_{eq}(\nu) = \left[\frac{\Omega_r(1+7\nu)}{\Omega_m(1+3\nu)+4\nu\Omega_r}\right]^{\frac{1}{1+3\nu}}\,.
\end{equation}
For $\nu=0$ we retrieve the standard result, as expected. Next we substitute the above expression for $a_{eq}(\nu)$ in Eq.\,\eqref{eq:E2RVM} in order to compute the normalized Hubble function for the RVM at equality, $E(a_{eq}(\nu)) $, and in this way we can  derive the ($\nu$-dependent) value of the wave number at equality, $k_{eq}(\nu)$. This allows us to compute the value of the argument $x(\nu)=k/(k_{eq}(\nu)\tilde{\Gamma})$  in  the transfer function (\ref{eq:BBKS}), which is obviously dependent on $\nu$ as well.   Let $T(x(\nu))$ be the corrected transfer function, and  let us call  $\Delta_T(x(\nu))= (T(x(\nu))-T_\CC(x))/T_\CC(x)$  the relative difference between the $\nu$-corrected transfer function and the standard transfer function $T_\CC(x)\equiv T(x(0))$. When $x\gg 1$ or, equivalently, when $k\gg k_{eq}$, the BBKS transfer function \eqref{eq:BBKS} can be approximated as follows:
\begin{equation}
T(x)\approx \frac{C\ln(1+Ax)}{x^2}\,,
\end{equation}
with $A=0.171$ and $C=(0.171\times 0.49)^{-1}$. The condition  $x\gg 1$ (i.e. $k\gg k_{eq}$) indeed holds in the most relevant integration region over the wave number, as  indicated in Fig. 2 (right) -- see the discussion below.  Needless to say, the exact form of $\Delta_T((x(\nu))$ is rather complicated for arbitrary $\nu$. However, $\nu$ is a very small parameter and therefore we can expand the above expressions linearly in $\nu$. After a detailed calculation, the final result can be cast in a truly compact form as follows (see \cite{GoSoPert} for an expanded discussion):
\begin{equation}\label{eq:DeltaT}
\Delta_T (x(\nu)\gg 1) =- \nu\left[7+6\ln\left(\frac{\Omega_m}{\Omega_r}\right)\right]+\mathcal{O}(\nu^2)\,.
\end{equation}
The most remarkable feature of this expression is that $\nu$ is enhanced by the big log factor  $\ln({\Omega_m}/{\Omega_r})={\cal O}(10)$. Exact numerical evaluation with the parameters of Table 1 yields $\Delta_T (x(\nu)\gg 1) =-8.8\%$.  This analytical estimate perfectly matches the asymptotic value of $\Delta_T$  that we have obtained from a direct numerical calculation  (cf. the left plot of Fig. 2). Therefore we have been able to explain the substantial correction (around $- 9\%$) undergone by the transfer function owing to the small parameter $\nu\sim 10^{-3}$, which effectively causes an enhancement proportional to $6\nu\ln({\Omega_m}/{\Omega_r}) \simeq 50\,\nu$.  However, this is not the end of the story, as this is not yet the final correction on $\sigma_8(z)$. The next step entails an integration over an expression that involves the transfer function--  see  Eq.\,\eqref{eq:s88general} --  and to incorporate the correction from  $\delta_m$; and, finally, also  the correction on the growth factor $f(z)$ so as to get the total effect on  $f(z)\sigma_8(z)$.  Of course $\delta_m$ and  $f(z)$ for the RVM  depend also on $\nu$ and are obtained from solving the modified perturbations equation (\ref{diffeqD}). The final correction after we subtract the result from the $\CC$CDM case is the following:
\be\label{eq:Dfs8}
\Delta_{f\sigma_8}(z;\nu)\equiv\,\frac{f(z;\nu)\sigma_8(z;\nu)-f(z)\sigma_8(z)}{f(z)\sigma_8(z)}\,.
\ee
Here $f(z)$  ensues from solving the standard perturbations equation (\ref{diffeqLCDM}). It turns out that the $\nu$-corrections on both $\delta_m(z;\nu)$ and  $f(z;\nu)$ are not comparatively significant to the above mentioned effects from the transfer function.  This follows from a close examination of the additional terms in the modified perturbations equation (\ref{diffeqD}) as compared to  (\ref{diffeqLCDM})  -- see \cite{GoSoPert} -- and can also be corroborated from  the explicit numerical test performed in Fig. 3. There we fix all parameters of the RVM to the central values of Table 1, and evaluate \eqref{eq:Dfs8}. We find e.g.  $\Delta_{f\sigma_8}=-6.3\%$ at $z=0$. This result  is, of course, not the real correction predicted by the RVM, which is around $-8\%$ (cf. Fig. 1 right) because the parameters of the $\CC$CDM are also fixed at the central values of the RVM, but with vanishing $\nu$.
With such strategy we can test the net effect of $\nu\neq 0$.  The obtained value can now be explained from the sum of the individual contributions from $\Delta_f(z)+\Delta_{\sigma_8}(z)$ at $z=0$, where from the structure of Eq.\,(\ref{eq:s88general}) we have $\Delta_{\sigma_8}(z)=\Delta_{\delta_m}(z)+\Delta_{\sqrt{I}}(z)$. Here $\Delta_{\sqrt{I}}$ is the variation of the $\sqrt{I}$ factor in (\ref{eq:s88general}) involving the integral $I$ of the product function $k^{n_s+2} T^{2}(k) W^2(kR_{8})$ over the relevant region of the the wave numbers, namely  $0.007\,{\rm Mpc}^{-1}\lesssim k \lesssim 0.3\,{\rm Mpc}^{-1}$ (cf. Fig. 2 right).  According to the left plot of this same figure and the aforementioned range of relevant $k$ values, we find  the relative differences in $\sqrt{I}$ to be approximately  $-5.5\%$  (the leading effect stemming from Eq.\,\eqref{eq:DeltaT} in this test). The sum with the previous (smaller) contributions accurately matches   $-6.3\%$ at $z=0$. The test holds good  for any  $z$, see Fig.\,3.


\begin{figure}
\begin{center}
\label{FigLSS2}
\includegraphics[width=3.2in]{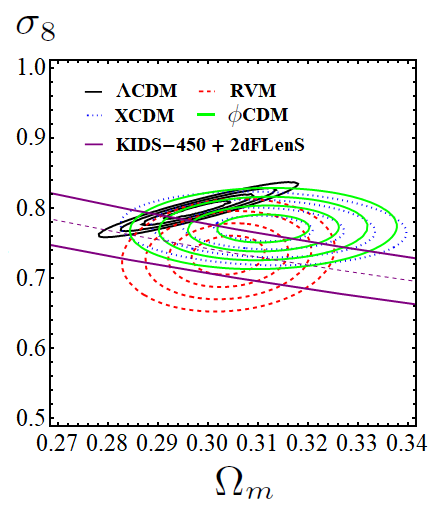}
\caption{{Likelihood contours  at $1\sigma$, $2\sigma$, $3\sigma$ and $4\sigma$ confidence level in the ($\Omega_m,\sigma_8)$ plane for the  $\Lambda$CDM, the XCDM, the original quintessence model with inverse-power potential\,\cite{PR88}  and the RVM}, together with the recent observational constraint provided by KiDS-450 and 2dFLenS collaborations \cite{Joudaki2017}, as extracted from weak gravitational lensing tomography and overlapping redshift-space galaxy clustering, $S_8=0.742\pm 0.035$ (purple curves). Shown is the allowed $1\sigma$ band for  $S_8$.  Dynamical DE is clearly favored, specially the RVM.
}
\end{center}
\end{figure}


The improvement in the description of the LSS data in the RVM does not only concern the $f(z)\sigma_8(z)$ data, but also some recent weak gravitational lensing constraints on $S_8\equiv \sigma_8(\Omega_m/0.3)^{0.5}$  (see e.g. \cite{Joudaki2017,Henning2017,Hildebrandt2017,Heymans2013}). This is crystal-clear from the plot of Fig. 4, where we show the contour lines in the ($\Omega_m,\sigma_8)$ plane for the $\CC$CDM, the XCDM and the RVM. {For completeness, we include also the contour lines for the quintessence model based on the Peebles \& Ratra potential $V(\phi)=({1}/{2})\kappa M_P^2\phi^{-\alpha}$ \cite{PR88}, see \cite{SoGoPeMPLA2017} for a recent analysis}.  The observational constraint in the same plane is provided by \cite{Joudaki2017},  $S_8=0.742\pm 0.035$, obtained from KiDS-450 and 2dFLenS collaborations from a joint analysis of weak gravitational lensing tomography and overlapping redshift-space galaxy clustering.  {This value of $S_8$ is $2.6\sigma$ below the one inferred in Planck's TT+lowP analysis \cite{Planck2016}}.  Our discussion and conclusions  remain virtually unchanged if we use e.g. the constraints from \cite{Hildebrandt2017} or \cite{Heymans2013}. Those by DES collaboration \cite{DES2017}, on the other hand, lie a bit higher in that plane, although they {are only mildly tensioned with the value of \cite{Joudaki2017}. In contrast, the aforementioned SPT measurements of the CMB EE and TE power spectra also lead to a lower value, $\sigma_8=0.770\pm 0.023$ \cite{Henning2017}. However, when these SPT data sets are combined with Planck's TT data the preferred $\sigma_8$ value shifts up to $\sigma_8=0.815\pm 0.014$, which is fully compatible with Planck's (and WMAP \cite{WMAP}) $\Lambda$CDM results, but  in conflict with the LSS data.}

From Fig. 4 we can see that a simple dynamical DE parametrization, such as the XCDM, already fits better the weak lensing data than the $\CC$CDM,  while the RVM best-fit value squarely hits the bull's-eye between the curves in purple from \cite{Joudaki2017}. We have to wait for future  measurements, but dynamical DE seems to be preferred right now by the available LSS and weak lensing data.

\section{Conclusions}

In this Letter we have been able to explain both  numerically and analytically the decisive improvement achieved  in the phenomenological  description of the  large scale structure formation data within the running vacuum model (RVM)  thanks to the small vacuum parameter $\nu\sim 10^{-3}$, which is responsible for the mild evolution of the  cosmic vacuum and is theoretically motivated in the context of  QFT in curved spacetime. The almost effortless and economical explanation offered by the RVM to relax  the $\sigma_8$-tension makes it a natural solution to such longstanding problem within the concordance model. In actual fact, the suggested solution invigorates the possible existence of significant signs of dynamical vacuum (or in general of dynamical DE) in the Universe, which were first reported in the recent works by  Sol\`a, G\'omez-Valent \& de Cruz P\'erez \cite{SoGoPeApJL2015,SoGoPeApJ2017} and  Zhao et al. \cite{GongBoZhao2017}.

\acknowledgments

\noindent We are partially supported by FPA2013-46570 (MICINN), Consolider CSD2007-00042, 2014-SGR-104 (Generalitat de
Catalunya) and  MDM-2014-0369 (ICCUB).

\end{document}